\documentclass[prd,aps,secnumarabic,showpacs]{revtex4}
\usepackage{amsmath}
\usepackage{amsfonts}
\usepackage{amssymb}
\usepackage{graphicx}
\def\part{\partial}

\begin{document}
\title{Comparison of $H\rightarrow\ell\bar{\ell}\gamma$ and $H\rightarrow\gamma\,Z, Z\rightarrow\ell\bar{\ell}$ including the ATLAS cuts}

\author{Duane A. Dicus$^a$\footnote{Email address: dicus@physics.utexas.edu},
Chung Kao$^b$\footnote{Email address: kao@physics.ou.edu},
Wayne W. Repko$^c$\footnote{Email address: repko@pa.msu.edu}}

\affiliation{
$^a$Department of Physics and Center for Particles and Fields,
University of Texas, Austin, TX 78712, USA \\
$^b$Homer L. Dodge Department of Physics, University of Oklahoma, Norman, OK 73019, USA \\
$^c$Department of Physics and Astronomy, Michigan State University, East Lansing, Michigan 48824, USA}
%

\date{\today}

\begin{abstract}
A precise comparison is made between the Dalitz decay $H\rightarrow\ell\bar{\ell}\gamma$ and the two body decay $H\rightarrow\gamma\,Z,Z\rightarrow\ell\bar{\ell}$ for electrons and for muons including experimental cuts appropriate for the ATLAS detector. The widths for these two processes differ by $8\%$ for electrons and $3\%$ for muons. Given that there remain QCD radiative corrections of this order that are not included, this suggests that the isolation of the Dalitz decay will be challenging.
\end{abstract}


\maketitle

{\bf Introduction} --
There have been several recent calculations of the width for Higgs decay into $\ell^{+}\ell^{-}\gamma$ where the $\ell$
are leptons \cite{DR},\cite{CQZ},\cite{GP},\cite{SCG}.
Each of these made rather crude approximations for the experimental cuts because the cuts are defined in the lab frame but the width is calculated in the rest frame where no axes are available to determine the rapidity or transverse momentum. To get around this difficulty what we do below is to define an average width where the average is over possible boosts.  The cuts can then be done on each term in the average. There are two reasons for doing this rather than just calculating the cross section for $p\,p\,\rightarrow\,H\,\rightarrow\,\gamma\ell\bar{\ell}+X$. One is that the cross section, because the Higgs can be on shell, requires a knowledge of the Higgs' total width. This method predicts the partial width without assuming a value for the total width.  The second is that the cross section requires QCD corrections to the production of $H$ and these can be uncertain by substantial factors. What we want to do here is a careful comparison between the direct decay and the decay through a $Z$ boson including the cuts precisely but without the complications of the full cross sections.

{\bf Method of Calculation} --
The width in the rest frame for $H\rightarrow\ell\bar{\ell}\gamma$ is given by
\begin{equation}\label{M}
\Gamma\,=\,\frac{\pi}{32}\frac{1}{(2\pi)^5}\frac{1}{2M_H}\int_0^{2\pi}d\phi \int^1_{-1}dz\int^{M_H^2}_{4m^2}ds\int^1_{-1}dz'\sqrt{1-\frac{4m^2}{s}}
\left(1-\frac{s}{M_H^2}\right)|M|^2\,,
\end{equation}
where $m$ is the lepton mass, $s$ is the invariant mass of the leptons, and $|M|^2$ is the square of the matrix element as given in Ref.\cite{ABDR}. The remaining variables are defined by the momenta of the particles as follows: The momenta of the individual leptons are defined in the their center of mass as
\begin{eqnarray}
p^{\mu}\,=\,\frac{\sqrt{s}}{2}(1,\beta\sqrt{1-z^2}\cos\phi,\beta\sqrt{1-z^2}\sin\phi,\beta\,z)\,\,,\\
p'^{\mu}\,=\,\frac{\sqrt{s}}{2}(1,-\beta\sqrt{1-z^2}\cos\phi,-\beta\sqrt{1-z^2}\sin\phi,-\beta\,z)\,,
\end{eqnarray}
where $\beta\,=\,\sqrt{1-4m^2/s}$. The momenta of the photon and lepton pair are defined in the Higgs rest frame as
\begin{eqnarray}
k^{\mu}\,&=&\,\frac{M_H^2-s}{2M_H}(1,\sqrt{1-z'^2},0,z')\,\,,\\
Q^{\mu}\,&=&\,\left(\frac{M_H^2+s}{2M_H},-\frac{M_H^2-s}{2M_H}\sqrt{1-z'^2},0,-\frac{M_H^2-s}{2M_H}z'\right)\,.
\end{eqnarray}
To determine the width in the Higgs rest frame we would boost $p^{\mu}$ and $p'^{\mu}$ to the frame of $Q^{\mu}$,
evaluate $|M|^2$ and evaluate the integrals in Eq.\,(\ref{M}).

\begin{figure}[h]\centering
\includegraphics[height=0.8in]{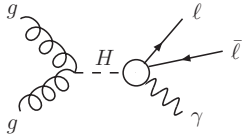}
\caption{The gluon fusion process for the Dalitz decay is shown. The circle indicates the full $H\to\ell\bar{\ell}\gamma$ loop amplitude given in Ref.\,\cite{ABDR}. The precise Higgs production cross section is not needed when calculating the average width $\tilde{\Gamma}$. \label{fusion}}
\end{figure}

To find the width in the lab frame let us define two ``lab'' momenta as
\begin{eqnarray}
p_1^{\mu}\,\equiv\,(x_1\,E_b,0,0,x_1\,E_b)\,\,,\\
p_2^{\mu}\,\equiv\,(x_2\,E_b,0,0,-x_2\,E_b)
\end{eqnarray}
where $E_b$ is the beam energy and $x_1,x_2$ are Bjorken scaling variables.  Now first boost $k^{\mu}$ and $Q^{\mu}$ into the lab frame given by $p_1+p_2$, then boost $p^{\mu}$ and $p'^{\mu}$ into the frame now given by the boosted $Q^{\mu}$. Then $|M|^2$ can be evaluated and $\Gamma$ can be obtained from  Eq.\,(\ref{M}). The fusion process is illustrated in Fig.\,(\ref{fusion}).

Now we have an axis defined by $p_1,p_2$ and, as we do the integrals, we can calculate things like rapidity and transverse momentum
and make cuts.  Of course each event will have a different set of $x_1,x_2$ so choosing particular values for these is not meaningful.  But what we can do is average over their values.  Define an average width as
\begin{equation}\label{R}
\widetilde{\Gamma}\,\equiv\,\frac{1}{N}\int_0^1\,dx_1\int_0^1\,dx_2\,g(x_1,M_H)g(x_2,M_H) \,\delta(\hat{s}-M_H^2)\,\Gamma
\end{equation}
where $\hat{s}\,=\,(p_1+p_2)^2\,=\,4E_b^2x_1x_2$, $\,\,g(x,M_H)$ are gluon distribution functions put in to weight the average, and $N$ is the same integral except without the $\Gamma$ and is included to normalize the average.

{\bf Experimental Cuts} --
The cuts for the  ATLAS detector are the following:
\begin{eqnarray}
m_{\ell\bar{\ell}}\,>\,81.2\,{\rm GeV}\,\,,\\
p_T^{\gamma}\,>\,15\,{\rm GeV}\,\,,\\
p_T^{\ell}\,>\,10\,{\rm GeV}\,\,,\\
|\eta_{\gamma}|\,<\,1.37\,\,\,{\rm or}\,\,\,\,1.52\,<\,|\eta_{\gamma}|\,<\,2.37\,\,,\\
|\eta_e|\,<\,2.47\,\,,\\
\Delta\,R_{\ell\gamma}\,>\,0.3
\end{eqnarray}
where $\eta$ is the pseudo-rapidity. The square of the invariant mass of the leptons, $m_{\ell\bar{\ell}}^2$, is called $s$ above so that cut can be made by changing the lower limit on the $s$ integral in Eq.\,(\ref{M}).   For muons the cuts are the same with the exception of $|\eta_\mu|\,<\,2.7$. 

\begin{figure}[h]\centering
\includegraphics[height=1.0in]{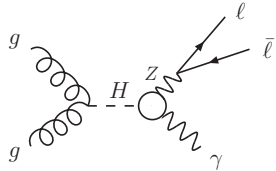}
\caption{The gluon fusion process for the sequential decay is shown. The  circle indicates the $W$ and $t$ triangle diagrams that are used to calculate the $H\to Z\gamma$ width \cite{BH,HHG}. Here, the $Z$ is understood to be on shell. The precise Higgs production cross section is not needed when calculating the average width $\widehat{\Gamma}$. \label{seq}}
\end{figure}

{\bf Results for electrons} --
Now we want to compare $\widetilde{\Gamma}$ with the width determined from the decay of $H\,\rightarrow\,\gamma\,Z$ multiplied by the branching ratio
for $Z\rightarrow\,\ell\bar{\ell}$. This is illustrated in Fig.\,(\ref{seq}). Of course we have to see the leptons so we must do the cuts here also. The decay width is a constant so in this case  we simply multiply by the fraction of phase space allowed by the cuts.   In other words consider
\begin{equation}
f_{PS}\,=\,\frac{\int^1_0dx_1\int^1_0\,dx_2\,g(x_1,M_H)g(x_2,M_H)\,\delta(\hat{s}-M_H^2)\, \int^{2\pi}_0d\phi\int^1_{-1}dz\int^1_{-1}dz'}{\int^1_0dx_1\int^1_0dx_2 \,g(x_1,M_H)g(x_2,M_H)\,\delta(\hat{s}-M_H^2)\,\,\,\,8\pi}
\end{equation}
and use the variables in the angular integrals to define $p^{\mu}, p'^{\mu}, k^{\mu}, Q^{\mu}$ just as above except that $s$ is replaced by $M_Z^2$ and we no longer integrate over $ds$ in Eq.\,(\ref{M}).  Define $p_1^{\mu}, p_2^{\mu}$ as above, do the boosts and check the cuts, setting $f_{PS}$ equal to zero when the cuts are not satisfied.  The result is the fraction of the phase space allowed by the cuts. 

The width for $H\rightarrow\,\gamma\,Z$, $\Gamma_{\gamma\,Z}$, is given in \cite{HHG}, and the branching ratio for $Z\rightarrow\ell\bar{\ell}$ is $B_{\ell\bar{\ell}}\,=\,0.03363$ for either electrons or muons. Thus the width for this, including the cuts above, is
\begin{equation}\label{R1}
\widehat{\Gamma}\,=\,\Gamma_{\gamma\,Z}\,B_{\ell\bar{\ell}}\,f_{PS} 
\end{equation}

Some results from Eqs.\,(\ref{R}) and (\ref{R1}) with electrons in the final state are given in Table I.
\begin{table}[h]
\begin{center}
\begin{tabular}{|c|c|c|c|} \hline
\,\,\,$E_{beam}(GeV)$\,\,\, & \,\,\,$M_H(GeV)$\,\,\, & \,\,\,\,\,\,\,$\widehat{\Gamma}(keV)$\,\,\,\,\,\,\, & \,\,\,\,\,\,\,$\widetilde{\Gamma}(keV)$\,\,\,\,\,\,\, \\  \hline\hline
3500. & 125. & 0.1307 & 0.1208  \\
  &  126. & 0.1440 & 0.1333  \\ 
4000. & 124. &   0.1153   &    0.1067 \\
  & 125. & 0.1276 & 0.1180  \\
  & 126. & 0.1406 & 0.1302  \\
  & 127. & 0.1545 & 0.1430  \\
7000. & 125. & 0.1155 & 0.1068  \\
  &  126. & 0.1272 & 0.1177  \\ \hline 
\end{tabular}
\end{center}
\caption[Table I]{$\widehat{\Gamma}$ is the width for $H\,\rightarrow\,Z\gamma$, $Z\,\rightarrow\,e^{+}e^{-}$.
$\widetilde{\Gamma}$ is the width for the Dalitz decay $H\,\rightarrow\,e^{+}e^{-}\gamma$}
\end{table}

The CTEQ6L1 distribution functions were used for the gluons. The values for $\widetilde{\Gamma}$ have a calculational error of $0.0005$. $B_{\ell\bar{\ell}}$ has an error of $0.00004$. There is a slight dependance on the beam energy because larger boosts make it harder to satisfy the $p_T$ and rapidity cuts. For example $f_{PS}$ varies from $59\%$ to $52\%$ for $M_H\,=\,125$\,GeV. But the result is that $\widehat{\Gamma}$ and
$\widetilde{\Gamma}$ differ by only $\sim8\%$ independent of beam energy or Higgs mass.

\begin{figure}[h]\centering
\includegraphics[height=0.9in]{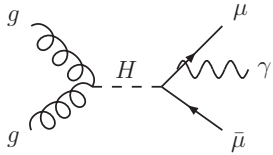}
\caption{The gluon fusion process for the direct muon coupling is shown. The precise Higgs production cross section is not needed when calculating the average width $\Gamma_2$. \label{muon}}
\end{figure}

{\bf Results for muons} --
Now consider muons in the final state.  $\widehat{\Gamma}$ is slightly larger because the allowed rapidity for muons is slightly bigger so the allowed fraction of phase space, $f_{PS}$, increases.  $\widetilde{\Gamma}$ now has two parts,
\begin{equation}
\widetilde{\Gamma}\,=\,\Gamma_1+\Gamma_2 
\end{equation}
where, as before, $\Gamma_1$ comes from the loop graphs in Eq.\,(\ref{R}) and $\Gamma_2$ is a tree contribution from the Higgs direct coupling to the lepton line, illustrated in Fig.\,(\ref{muon}).  

The interference between the loop and tree graphs is negligible \cite{AR} and the average of the tree part is
\begin{eqnarray}
\Gamma_2\,&=&\,\frac{1}{N}\int_0^1\,dx_1\int_0^1\,dx_2\,g(x_1,M_H)g(x_2,M_H)\delta(\hat{s}-M_H^2) \nonumber \\
&&\int_0^{2\pi}d\phi\int^1_{-1}dz\int^{M_H^2}_{4m^2}\,ds\int_{-1}^1\,dz'\sqrt{1-\frac{4m^2}{s}}\left(1-\frac{s}{M_H^2}\right) \frac{\alpha^2}{128\pi^2\,M_H\sin^2\theta_W}\left(\frac{m}{M_W}\right)^2\frac{s^2+M_H^4}{tu}\,\,\label{Tree}
\end{eqnarray}  
where $u\,=\,2p\,\cdot\,k+m^2$ and $t\,=\,M_H^2-s-u+2m^2$. As before we use $\phi$, $z$, $s$, and $z'$ to construct the momenta $p^{\mu}, p'^{\mu}, k^{\mu}$ and $Q^{\mu}$ and $x_1, x_2$ to construct $p_1^{\mu}$ and  $p_2^{\mu}$,  boost $Q$ and $k$ to the frame of $p_1+p_2$ and $p$ and $p'$ to the frame of $Q$.  Then we use these boosted momenta to construct the $p_T$, $\eta$ and $\Delta\,R$, check the cuts, and set (\ref{Tree}) equal to zero if any cut is not satisfied.   Again $N$ is the integral over $x_1$ and $x_2$ of the distribution functions and the delta function. Numerically $\Gamma_2$ turns out to be about $5\%$ of $\Gamma_1$.

The results for muons are given in Table II. The addition of $\Gamma_2$, which has nothing to do with the $Z$ boson, reduces the difference between $\widehat{\Gamma}$ and $\widetilde{\Gamma}$
to about $3\%$.
\begin{table}[h]
\begin{center}
\begin{tabular}{|c|c|c|c|c|c|}\hline
\,\,\,$E_{beam}(GeV)$\,\,\,&\,\,\,$M_H(GeV)$\,\,\,&\,\,\,\,\,$\widehat{\Gamma}(keV)$\,\,\,\,\,&\,\,\,\,\, $\Gamma_1(keV)$\,\,\,\,\,&\,\,\,\,\,$10^2\,\Gamma_2(keV)$\,\,\,\,\,&\,\,\,\,\,$\widetilde{\Gamma}(keV)$\,\,\,\,\,  \\ \hline\hline
3500. & 125. & 0.1375 & 0.1270 & 0.6656 & 0.1337  \\
      & 126. & 0.1515 & 0.1400 & 0.6891 & 0.1469  \\
4000. & 124. & 0.1216 & 0.1126 & 0.6272 & 0.1189  \\
      & 125. & 0.1346 & 0.1244 & 0.6516 & 0.1309  \\
      & 126. & 0.1484 & 0.1373 & 0.6740 & 0.1440  \\
      & 127. & 0.1630 & 0.1510 & 0.6996 & 0.1580  \\
 7000.& 125. & 0.1225 & 0.1132 & 0.5968 & 0.1192  \\
      & 126. & 0.1349 & 0.1248 & 0.6167 & 0.1310  \\ \hline
\end{tabular}
\end{center}
\caption{$\widehat{\Gamma}$ is the width for $H\,\rightarrow\,Z\gamma$, $Z\,\rightarrow\,\mu^{+}\mu^{-}$.  
$\widetilde{\Gamma}\,=\,\Gamma_1+\Gamma_2$ is the width for the Dalitz decay $H\,\rightarrow\,\mu^{+}\mu^{-}\gamma$.}
\end{table}

{\bf Summary} -- 
We have tried to include precisely the experimental cuts of the ATLAS experiment in the width for $H\,\rightarrow\,\ell\bar{\ell}\gamma$.
The detailed results are given in Table I and Table II.
For both electrons and muons the two ways for the decay to take place, Dalitz decay given by $\widetilde{\Gamma}$ and decay through the $Z$ as given by 
$\widehat{\Gamma}$, have widths
that differ by only a few per cent.  Both processes will have radiative corrections that have not been included here and those corrections should be
a few per cent in magnitude.  Thus, at this stage, the widths determined in the two ways must be considered equal.

{\bf Acknowledgements} -- It is our pleasure to thank the  ATLAS group whose questions led us to do this work, Giovanni Marchiori, Aiden Sean Randle-Conde,
and especially Efstathes Paganis.  D.A.D. was supported in part by the U.S.Department of Energy under Award No. DE-FG02-12ER41830. C.K. was supported in part by U.S.Department of Energy under Award No. DE-FG02-13ER41979. W.W.R. was supported in part by the National Science Foundation under Grant No. PHY 1068020.

\end{document}